\documentclass[usenatbib]{mnras}

\usepackage{amsmath}
\usepackage{siunitx}
\usepackage{xfrac}
\usepackage{natbib}
\usepackage{xcolor}
\usepackage{graphicx}
\usepackage{booktabs}
\usepackage{floatrow}
\usepackage{caption}
\usepackage{csquotes}
\usepackage{hyperref}
\usepackage{ulem} 
\usepackage[varg]{txfonts} 

\usepackage{hyperref}            
            
\hypersetup{pdfpagemode = {UseNone},
            pdftitle = {Planet composition},
            pdfcreator = {\LaTeX},
            pdfproducer = {pdfeTeX-0.\the\pdftexversion\pdftexrevision},
            pdfauthor = {AP, AZ},
            pdfsubject = {},
            pdfview = {FitH},
            pdfstartview = {FitH},
            colorlinks = {true},
            linkcolor = [rgb]{0,0.35,0.7},
            citecolor = [rgb]{0,0.35,0.7},
            filecolor = [rgb]{0.61,0,0},
            urlcolor = [rgb]{0.61,0,0},
           }
           
\usepackage{bm}

\newcommand*\dd[1]{\,\mathrm{d}#1}

\newcommand{\tensorGR}[1]{\overline{\bm{{#1}}}}
\newcommand{\DP}[2]{\frac{\partial{#1}}{\partial{#2}}}

\newcommand{\eb}{e_\mathrm{bin}}
\newcommand{\ab}{a_\mathrm{bin}}
\newcommand{\acav}{a_\mathrm{cav}}
\newcommand{\ecav}{e_\mathrm{cav}}
\newcommand{\bin}{_\mathrm{bin}}

\newcommand{\G}{\text{G}}
\newcommand{\Mco}{M_\text{co}}
\newcommand{\Msun}{\text{M}_\odot}
\newcommand{\Rgas}{\mathcal{R}}
\newcommand{\cs}{c_\text{s}}

\newcommand{\OmegaK}{\Omega_\text{K}}

\newcommand{\cv}{c_\text{v}}
\newcommand{\sigmaSB}{\sigma_\text{SB}}
\newcommand{\vel}{\bm{u}}

\colorlet{darkgreen}{green!60!black}

\newcommand{\Tb}{T_{\mathrm{bin}}}

\newcommand{\lrad}{l_\text{rad}}
\newcommand{\pluto}{\texttt{PLUTO}}

\newcommand{\beq}{\begin{equation}}
\newcommand{\eeq}{\end{equation}}

\title[Circumbinary cavities at different scales]{Sculpting protoplanetary discs --- modelling circumbinary cavities at observable scales with radiation hydrodynamics}

\author[Penzlin et al.]{
Anna~B.T. Penzlin,$^{1,2}$\thanks{E-mail: a.penzlin@imperial.ac.uk}
Alexandros Ziampras,$^{2,3,4}$\thanks{E-mail: a.ziampras@lmu.de}
Nicolas T. Kurtovic,$^{5}$
\newauthor
Marcelo Barraza-Alfaro,$^{6}$
Paola Pinilla$^{7}$
\\
$^{1}$Astrophysics Group, Department of Physics, Imperial College London, Prince Consort Rd, London, SW7 2AZ, UK \\
$^{2}$Ludwig-Maximilians-Universit{\"a}t M{\"u}nchen, Universit{\"a}ts-Sternwarte, Scheinerstr.~1, 81679 M{\"u}nchen, Germany\\
$^{3}$Astronomy Unit, School of Physics and Astronomy, Queen Mary University of London, London E1 4NS, UK\\
$^{4}$Max-Planck-Institut f{\"u}r Astronomie, K{\"o}nigstuhl 17, 69117 Heidelberg, Germany\\
$^{5}$Max-Planck Institut für Extraterrestrische Physik (MPE), Giessenbachstr. 1, 85748, Garching, Germany\\
$^{6}$ Department of Earth, Atmospheric, and Planetary Sciences, Massachusetts Institute of Technology, Cambridge, MA 02139, USA\\
$^{7}$ Mullard Space Science Laboratory, University College London, Holmbury St Mary, Dorking, Surrey RH5 6NT, UK\\
}

\date{Accepted XXX. Received YYY; in original form ZZZ}

\pubyear{2024}

\begin{document}
\label{firstpage}
\pagerange{\pageref{firstpage}--\pageref{lastpage}}
\maketitle

\begin{abstract}
{
Observations of circumbinary discs reveal inner cavities, with their shape and size varying strongly between different systems. The structure of the cavity is determined by the complex interplay between spirals induced by tidal forcing from the binary and the viscous and radiative damping of the spirals at the cavity edge. To fully understand what determines the properties of observed cavities, it is therefore necessary to capture the effect of radiative processes in modelling. To this end, we run 27 simulations of circumbinary discs in 2D using the \pluto{} code. These simulations include various size scales, binary eccentricities and thermodynamic models. 
We find that the diverse cavity shapes are a natural outcome of the radially-varying cooling timescale, as different radiative processes mediate cooling at different disc size regimes. For binaries with separation of a few au, where the cooling timescale is comparable to the orbital timescale at the cavity edge, we recover much more circular cavities than for quickly- or slowly-cooling discs. Our results show that the cavity structure around several binary systems such as Cs~Cha and GG~Tau can be explained with one physical model, and highlight the importance of radiative cooling in modelling the dynamical evolution of circumbinary discs.
}
\end{abstract}


\section{Introduction}\label{sec:intro}


One of the first ever resolved images of protoplanetary discs was the image of the disc around GG~Tau~A, revealing a larger inner cavity ($\sim 180$ au) carved by the inner multi-star motion \citep{1999Guilloteau}. With ALMA and VLT (e.g. with instruments like SPHERE and GRAVITY)
 our capability to observe discs has improved dramatically, pushing the resolution limit to inner cavities of $\sim10$ au with ALMA (Atacama Large Millimeter/submillimeter Array) and detecting orbiting disc material at sub-au orbits with the VLT (Very Large Telescope) through instruments like SPHERE and GRAVITY. 
The size and shape of the circumbinary disc changes when considering different disc scales, however. While a large, eccentric cavity and extended spirals within the inner disc have been observed for GG~Tau~A \citep{2020Keppler} and HD~142527 \citep{2021Hunziker}, the behaviour changes to circular, smaller cavities with only axis-symmetric, narrow inner disc features like in Cs~Cha \citep{2022Kurtovic}, and again to asymmetric features around spectroscopically close binary systems \citep{2020Kluska}.

Current models of circumbinary discs have not yet established a clear link between the physical separation of the binary and the structure of the disc and its inner cavity, as most models are either parametrized with a through a scale-free, locally isothermal model \citep[e.g.][]{2017Thun,2020Ragusa,2022Dittmann,2024Penzlin}, or focus on one size regime like the circumbinary Kepler-planet hosts with separation $<1$~au \citep[e.g.][]{2019Kley,2021Pierens}.
While the very large disc like GG~Tau and HD~142527 with cavities of $>100$~au might be optically thin enough to cool on nearly locally isothermal timescales.
The models leave the intermediate regime of small circular cavities unaddressed, and requires invoking additional processes to explain their occurrence. One such mechanism is dust--gas interaction, which can dampen eccentricity forcing \citep{2022Coleman} but necessitates dust-to-gas ratios and dust sizes well above the stability threshold for the streaming instability \citep{2005Youdin,2007Johansen} to fully circularize the cavity edge. Alternatively, embedded massive planets can carve out a circular gap, eliminating the eccentric cavity edge \citep{2022Kurtovic}, though their presence remains unverifiable and raises the question of how such planets reached the cavity edge in the first place. \cite{2022Fitzmaurice} raises the problem destabilisation of small planets through the resonant interaction with the binary and \cite{2008Pierens} finds that giant planets around binaries can be scattered when migration leads them too close to the binary. Therefore, to explain the variation in cavity structures across different scales, there is strong motivation to identify a simpler model based on the intrinsic properties of disc hydrodynamics.

One important behaviour changing with the scale of the system is
the heating of the circumbinary disc through viscous, shock, or stellar irradiation, and its cooling via thermal emission and radiative diffusion. 
As a result, small-scale discs are typically optically thick and therefore cool very slowly, while at large radial scales the disc is so optically thin that cooling is limited by the inefficient radiative properties of dust and their weak coupling to the gas. At intermediate ranges of $\sim$10--50\,au in the disc, cooling is expected to be sufficiently efficient to operate on dynamical timescales \citep{bae-etal-2021}. This behaviour opens the possibility to interpret the different cavity properties as the natural outcome of the radially-varying cooling timescale, provided the latter affects the cavity-opening process.

To that end, \cite{2022Sudarshan} have shown how radiative damping of the spiral arms launched by the central binary can affect the size and shape of the circumbinary cavity in general, further noting the compatibility of their model with the Kepler-planet like circumbinary systems. They found that thermal relaxation on the orbital time scales leads to a significantly weaker excitation of eccentricity in the binary cavity. To explain their findings they then drew a parallel to work in the planet--disc interaction context \cite{2020Miranda_I, 2020Miranda_II, 2020ZhangZhu, 2020Alex_II}, which has shown that radiative damping of planet-driven spiral wakes results in angular momentum deposition and therefore gap opening only very close to the planet's orbital radius when the cooling time is comparable to the orbital time. In both cases, the result is narrow cavities around the perturber's orbit, and little to no structure elsewhere in the disc.


In this work, we use numerical hydrodynamics simulations to build on the results of \citet{2022Sudarshan}, adopting a refined model of radiation thermodynamics with a treatment of various radiative processes that determine the thermal structure of the disc. Our approach captures the radial dependence of the cooling timescale in a protoplanetary disc, allowing us to investigate the properties of circumbinary cavities at different radial scales that often align with those observable with ALMA or VLT. We then establish connections to observed circumbinary systems to test our findings and the applicability of our model.

In Sect.~\ref{sec:theory} we introduce our physical framework and discuss the radiative terms included in our models. We describe our numerical setup in Sect.~\ref{sec:setup}, and provide predictions for the radial profile of the cooling timescale in Sect.~\ref{sec:thermal-models}. We present the results of our hydrodynamical simulations and synthetic observations in Sects.~\ref{sec:results-shape}~and~\ref{sec:synthetic}, respectively. We discuss our findings in Sect.~\ref{sec:discuss}, and conclude in Sect.~\ref{sec:conclusion}.

\section{Hydro- and thermodynamics in accretion discs}
\label{sec:theory}

We will build on the vertically integrated, two-dimensional (2D) models presented in \cite{2019Kley}, using a GPU-parallelised version of the \pluto{} code \citep{2007Pluto, 2017Thun}. We therefore adopt a similar 2D setup, where we assume the binary and its surrounding disc to be co-planar. 

\subsection{Hydrodynamics}

To compute the dynamics of the system, the code solves the following set of vertically integrated hydrodynamic equations:
\begin{equation}
	\label{eq:navier-stokes}
	\begin{split}
	\DP{\Sigma}{t} & + \nabla\cdot(\Sigma\vel) = 0, \\
	\DP{\Sigma\vel}{t}& + \nabla\cdot\left(\Sigma\vel\otimes\vel + P \textbf{I} - \tensorGR{\Pi}\right)= \Sigma \bm{g}, \\
	\DP{E}{t}& + \nabla\cdot\left((E + P - \tensorGR{\Pi})\vel\right)=\Sigma\vel\cdot\bm{g} + Q. \\
	\end{split}
\end{equation}
Here $\Sigma$ is the gas surface density, $\vel$ is the velocity vector, $P$ is the vertically integrated pressure which are the variables evolved through the set of equations.
Through these quantities the total energy density $E=e + \Sigma u^2/2$ can be determined using the internal energy density for an ideal gas $e=P/(\gamma-1)$ with adiabatic index $\gamma=7/5$. $\bm{g}$ is the gravitational acceleration from both binary components given by
%
\begin{equation}
    \label{eq:gravity-acceleration}
    \bm{g} = \sum\limits_i -\frac{\G M_i}{s^3_i} \bm{s}_i,\qquad \bm{s}_i = \bm{R} - \bm{R}_i
\end{equation}
where $\bm{R}$ is the position vector of a gas parcel and $\bm{s}_i$ its distance to the star with index $i$ and mass $M_i$. For the viscosity $\tensorGR{\Pi}$ we assume a shear viscosity $\nu = \alpha \cs^2/\Omega$ \citep{1973alpha} with $\alpha=10^{-4}$.

The sound speeds $\cs$ relate to the pressure and temperature of the gas through
\begin{equation}
\cs/\sqrt{\gamma}=\sqrt{P/\Sigma}=H\Omega= \sqrt{T  \frac{\Rgas}{\mu}},\label{eq:cs}
\end{equation}
where $T$ is the gas temperature, $\Rgas$ is the gas constant and $\mu=2.35$ is the mean molecular weight of the gas. Thereby, the current temperature can be evaluated through pressure and density. In case of isothermal models the adiabatic index is considered to be $\gamma=1$. The binary potential locally alters the pressure--gravity balance that determines the local angular velocity $\Omega$ by:
\begin{equation}
\Omega = \sqrt{\sum_i \frac{\G M_i}{R_i^3}}.
\end{equation}

Finally, the heating and cooling source terms $Q$ are the sum of irradiation, viscous heating and radiative cooling. These are determine by the current state, before the next numeric step. We describe these terms further in the paragraphs below. 

\subsection{Locally isothermal assumption}

The locally isothermal assumption uses a fixed temperature profile ($\partial T/\partial t=0$) in which the temperature scales with the radially-dependent aspect ratio $h = H/R$. As we later use a irradiation heating source, we use a flaring index of 2/7 \citep[i.e., $h(R)= h_0 R^{2/7}$, see also][]{1997ChiangGoldreich}. 
The temperature can then be calculated as:
\begin{equation}
	\label{eq:temperature-aspect}
	T = \frac{\mu\G \Mco}{\Rgas} \frac{h^2}{R}, \qquad \Mco = \sum\limits_i M_i.
\end{equation}

To set the initial condition for the simulations and the thermal profile for the isothermal case we, will scale the value of the $h_0$ constant in this equation, such that it creates a consistent disc profile across all disc sizes as shown in Fig. \ref{fig:1d_hr}. When comparing the radiative and locally isothermal model, this simple locally isothermal profile based on irradiation is a good approximation and, hence, a sensible initial condition for the discs.

\subsection{Heating sources}

At small radial scales, where we are dealing with the dense inner disc, viscous heating becomes important. This is given in our non-isothermal models by \citep{1978Tassoul}
\begin{equation}\label{eq:Q_visc}
    Q_\text{visc} = \frac{1}{2\nu\Sigma} \text{Tr}\left(\tensorGR{\Pi}\right)\approx \frac{9}{4}\nu\Sigma\Omega^2.
\end{equation}

We also treat stellar heating by considering the irradiation flux from each star at distance $s_i = |\bm{R}-\bm{R}_i|$. The heating rate is then \citep{2004MenouGoodman}
%
\begin{equation}
    Q_\text{irr} = \sum\limits_i 2\frac{L_i}{4\pi s_i^2}(1-\varepsilon)\left(\frac{\mathrm{d}\log H_\text{s}}{\mathrm{d}\log R} - 1\right)\frac{H_\mathrm{s}}{R}\tau_\mathrm{eff}^{-1} \label{eq:Q_irr},
\end{equation}
where $\tau_\mathrm{eff}$ is an effective optical depth (see below), $\varepsilon=0.5$ is the disc albedo and $H_\mathrm{s} = 4H$ is the height of the disc optical surface \citep{1997ChiangGoldreich}. We do not consider the effect of shadowing, and fix $\frac{\mathrm{d}\log H_\text{s}}{\mathrm{d}\log R} = 9/7$. 

\subsection{Radiative cooling}

The disc cools through thermal emission via its surfaces at a rate $Q_\text{cool}$ given by
\begin{equation}\label{eq:Q_cool}
	Q_\text{cool} = -2 \sigmaSB \frac{T^4}{\tau_\mathrm{eff}},
\end{equation}
where $\sigmaSB$ is the Stefan-Boltzmann constant. The effective optical depth $\tau_\text{eff}$ can be calculated with the emission optical depth $\tau$ following \cite{1990Hubeny} as
\begin{equation}
\tau_\mathrm{eff} = \frac{3\tau}{8} + \frac{\sqrt{3}}{4} + \frac{1}{4\tau},\quad \tau=\int\limits_{0}^\infty\kappa\rho\dd z\approx\frac{c_1}{\sqrt{2\pi}}\kappa\Sigma,
\end{equation}
where $c_1=0.5$ is a constant introduced to match the results of three-dimensional simulations \citep{2012Muller}. We use the disjoint power-law opacity model of \citet{1985Lin} \citep[see also][]{2012Muller}, and further assume that the Rosseland and Planck mean opacities are equal such that $\kappa_\text{R} = \kappa_\text{P}=\kappa$.

\subsection{In-plane cooling}

While the vertical cooling channel discussed in the previous paragraph is largely responsible for setting the disc temperature profile, a non-negligible radiative flux flows through the disc plane as well. Radiative diffusion through the disc midplane can act as an efficient cooling channel, typically 3--4 times more efficient in removing temperature perturbations than surface emission \citep{2020Miranda_II,ziampras-etal-2023} and much more prominent in the presence of strong temperature gradients \citep[e.g.,][]{ziampras-etal-2025}. This can be expressed for example in the flux-limited diffusion approximation \citep{1981Levermore}
\begin{equation}
\label{eq:Q_rad}
Q_\text{rad} = \sqrt{2\pi} H \nabla \cdot \left( \lambda \frac{4\sigma_\text{SB}}{\kappa_\text{R}\rho_\text{mid}}\nabla T^4 \right),
\end{equation}
where $\rho_\text{mid}=\frac{1}{\sqrt{2\pi}}\frac{\Sigma}{H}$ is the midplane volume density and $\lambda$ is a flux limiter that handles the transition between the optically thick, diffusive limit and the optically thin, free-streaming limit. In our models, we implement cooling due to this radiative flux by relaxing the temperature to its initial profile over a cooling timescale $t_\text{cool} = \beta\OmegaK^{-1}$ given by \citep[e.g.][]{2017Flock}

\begin{equation}\label{eq:beta_rad}
	\beta_\text{mid} = \frac{\Omega_\text{K}}{\eta}\left(H^2 + \frac{\lrad^2}{3}\right),\quad\eta=\frac{16\sigma_\text{SB} T^3}{3\kappa\rho^2\cv},\quad \lrad = \frac{1}{\kappa\rho},
\end{equation}
where $\eta$ is the radiative diffusion coefficient, $\cv$ is the heat capacity for constant volume, and $\lrad$ is the photon mean free path. While this approach does not capture the diffusive component of the in-plane radiative flux,
it handles the transition from optically thick (where the characteristic length scale is $l_\text{c}\sim H$) to optically thin regions ($l_\text{c} \sim \lrad$) similarly to Eq.~\eqref{eq:Q_rad} by requiring that cooling is limited by radiative diffusion and emissivity in the optically thick and thin limits, respectively. The relaxation source term to a reference profile $T_0(R)$ is then given by \citep[e.g.][]{2001Gammie}
\begin{equation}
    \label{eq:beta-cooling}   
	\DP{e}{t} = Q_\text{relax} = -4\Sigma\cv\frac{T-T_0}{\beta_\text{mid}}\OmegaK,
\end{equation}
with the factor of 4 motivated by linear analysis in order to match calculations with a self-consistent treatment of radiative terms \citep{dullemond-etal-2022,ziampras-etal-2023}.

\section{Setup}\label{sec:setup}

In this section, we describe our numerical setup and initial conditions. We then list the procedure we followed to maintain a fair comparison between radiative and locally isothermal models.

\subsection{Numerical setup}
\label{sub:numerics}

We use the \pluto{} astrophysical code \citep{2007Pluto} in a vertically integrated, cylindrical polar geometry $\{R,\phi\}$. For a given binary separation $\ab$ our grid extends radially between $R\in[1,40]\,\ab$ with a logarithmic spacing and covers the full $2\pi$ in the azimuthal direction with linear spacing. We use a fully second-order accurate scheme with piecewise linear reconstruction and RK2 time stepping along with the HLL solver \citep{Toro} for stability and the \texttt{CHAR\_LIMITING} flag to improve accuracy by reconstruction on the characteristic variables. The orbital evolution of the binary is integrated in time using the N-body scheme detailed in \citet{2017Thun}.

Similar to \citet{2019Kley} and \citet{2022Sudarshan}, our radial boundary maintains a strict outflow condition at the inner disc edge, while all quantities are reset to their initial values at the outer boundary. This combination of boundary conditions with the radial extent of the domain is motivated by the analysis in \citet{2022Penzlin}, who found that this configuration is appropriate to resolve cavity dynamics while maintaining a steady disc profile.

\subsection{Physical conditions}

In this study, we use a low but non-zero turbulent viscosity with $\alpha=10^{-4}$, which can still contribute substantial viscous heating, especially at small scales.
Our choice of $\alpha$ is motivated by three-dimensional simulations, which have shown that binary-induced parametric instabilities are likely to introduce turbulence of this order \citep{2020Pierens}. This is consistent with the settling and radial width of features in recent dust continuum observations \citep{2022Villenave, 2018DSHARP_VI}, which have indicated that level of turbulence are of the order of $\alpha=10^{-5}$--$10^{-3}$ in typical protoplanetary discs.

As numerous studies \citep[e.g.][]{2017Thun,2020Ragusa,2023siwek} already showed that the binary eccentricity $\eb$ changes the shape of the disc, we will consider 3 values of eccentricity $\eb=[10^{-2}, 0.15, 0.3]$. 
Previous locally isothermal models \citep{2018Thun} showed high disc eccentricities for $\eb=0$ and $\eb\geq0.3$ and minimal values for $\eb\approx0.15$.
To put an intermediate-sized disc at the centre of the investigation, we base the stellar properties on the case of Cs~Cha with a combined binary mass of $\Mco=1.9~\Msun$. The binary stars are of equal mass with a luminosity of $0.745~L_\odot$ each and the orbits of the binary do not receive any feedback from the disc.

To demonstrate the role of the spatially varying cooling timescale we chose three different binary separations $\ab=\{1, 5, 25\}\,$au, corresponding to the optically thick, marginally thick, and optically thin disc regimes, respectively. The smallest value corresponds in size to the initial host system of observed circumbinary planets. The middle value is the estimated separation of the binary stars in Cs~Cha \citep{2007Guenther} and close to the separation in HD~100546 \citep{2024Stolker}, and the largest scale is comparable with large disc systems like GG~Tau \citep{2020Keppler} or HD~142527 \citep{2024Stolker}.

For these different cases, we initialise the surface density as:
\begin{equation}
\Sigma(R) = \Sigma_\text{ref} \left(\frac{R}{\ab}\right)^{-15/14},
\end{equation}
with a reference surface density at $1~\ab$ as $\Sigma_\text{ref}=[1.01, 4.5, 20.1]\times 10^{-4} M\bin \ab^{-2}$ for the three size scales, or $[1700, 1363, 1086]~\text{g}/\text{cm}^2$ at 1\,au. The surface density exponent is reproducing a steady disc with a flaring index of 2/7 following the analytical expectation of the irradiation profile given by Eq.~\eqref{eq:Q_irr}. A summary of the parameters is given in Table \ref{tab:parameters}.

The locally isothermal models are then designed to match the density distribution and equilibrium aspect ratios as the models with temperature evolution. Simulations are done in 3 steps:

\begin{itemize}
\item Radiative models with viscous heating, stellar irradiation and radiative cooling (Eqs.~\eqref{eq:Q_visc},~\eqref{eq:Q_irr}, \eqref{eq:Q_cool}) are carried out to allow the disc to reach a quasi-equilibrium state and compute realistic surface temperature estimates.
\item Corresponding locally isothermal models using the above temperature profiles (approximated as power-laws) are executed in parallel for comparison.
\item In-plane cooling (Eq.~\eqref{eq:Q_rad},~\eqref{eq:beta_rad}) is included into the radiative models after 10\,000 binary orbits with $T_0$ determined from above, and simulations are continued for 30\,000 orbits.
\end{itemize}

\begin{figure}
	\centering
    \includegraphics[width=1.02\columnwidth]{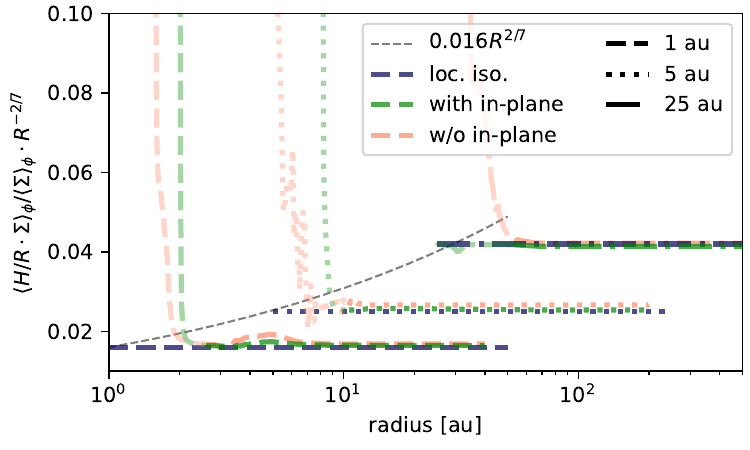}
	\caption{Mass-weighted azimuthally averaged aspect ratios against radius. The scale is readjusted to irradiative flaring and a constant line represents a $R^{2/7}$ radial flaring.
    The dark blue lines represent locally isothermal models, green lines models that include all heating terms, radiative cooling through the surface and in-plane cooling and light red are models with heating and cooling but without in-plane cooling. Different line style represent different separations (1~au dashed; 5~au potted; 25~au dash-potted). Regions inside the cavity are in lighter colours.}
	\label{fig:1d_hr}
\end{figure}

Following these steps all discs are directly comparable in terms of thermal structure, with similar scale heights as shown in Figure~\ref{fig:1d_hr}, but with the core difference being the treatment of radiative effects. 
The pale sections in the figure represent the averaged conditions within the cavity, where densities are more than three orders of magnitude lower than within the disc. There, spiral shocks can easily create temperature peaks following Eq.~\ref{eq:cs}, which carry insignificant thermal energy but lead to spikes in the scale height when averaged over.
All models are run for 30\,000 binary orbits in total.

\begin{table}
\begin{tabular}{|c|c|c|c|c|c|c|}
\hline 
$\alpha$ & $M_{1,2}$ [$\Msun$] &$L_{1,2}$ [$L_\odot$] &  $\eb$ & $\ab$[au] & $h_0$ [\%] & $f$\\ 
\hline 
$10^{-4}$ & 0.95 & 0.745 & 0.01 & 1& 1.6& 2/7 \\ 
 & & & 0.15& 5 & 2.5 & \\ 
 & & & 0.3 & 25 & 4.2& \\ 
\hline 
\end{tabular}
\caption{Model parameters. The same viscous $\alpha$ and the stellar masses $M_{1,2}$ and luminosities $L_{1,2}$ in solar units are used in all simulations, while the binary eccentricity $\eb$ and the binary semi-major axis $\ab$ are varied independently. The aspect ratio of the isothermal and initial aspect ratio of the evolving models $h_0$ dependence on the value of $\ab$ and flares with the exponent $f$.}
\label{tab:parameters}
\end{table}

\section{Thermal models of circumbinary discs} \label{sec:thermal-models}

\begin{figure}
	\centering
    \includegraphics[width=250pt]{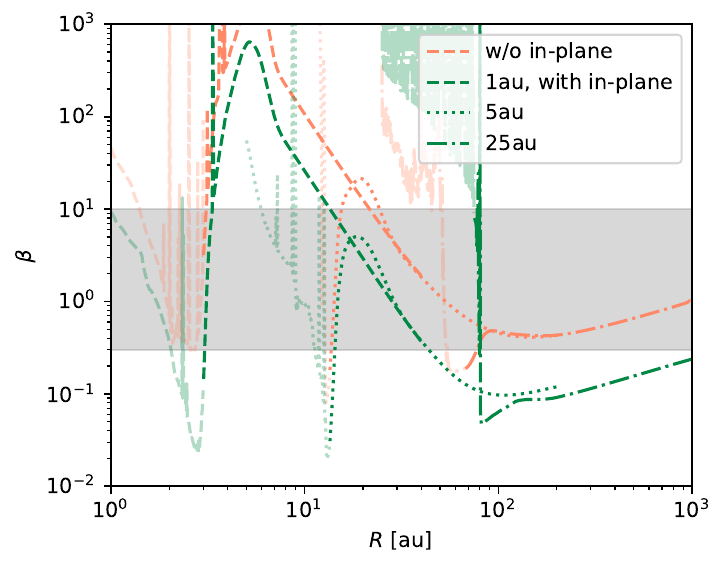}
	\caption{Azimuthally averaged cooling timescale profiles. The green lines represent models and simulations with in-plane cooling, the light red lines without in-plane cooling. Line styles indicate model with a binary separation of ($\ab = 1$~au: dashed; $\ab = 5$~au: dotted; $\ab = 25$~au: dashed-dotted). Inside the cavity, lines turn pale.}
	\label{fig:1d_beta}
\end{figure}

The heating and cooling budget of the disc depends on the flaring angle and the temperature- and density-dependent opacity, both of which are scale-dependent in the circumbinary disc.
In Figure \ref{fig:1d_beta} we show the contributions of heating and cooling and the resulting cooling timescale for the azimuthally averaged models. The latter is defined as simply
\begin{equation}
    \label{eq:beta-surf}
    \beta_\mathrm{surf} = \frac{e}{|Q_\mathrm{cool}|}\OmegaK
\end{equation}
in models where the in-plane cooling contribution is ignored (tagged ``w/o in-plane''), and
\begin{equation}
    \label{eq:beta-tot}
    \beta_\mathrm{tot}^{-1} = \beta^{-1}_\mathrm{surf} + \beta^{-1}_\mathrm{mid}
\end{equation}
in models where both surface and in-plane cooling are considered (tagged ``with in-plane'', see also Eq.~\eqref{eq:beta_rad}). This formulation \citep{2020Miranda_II} has been shown to match with radiative models very well \citep{ziampras-etal-2023}, accurately representing the overall cooling efficiency of the disc.

The long cooling time between 3 and 7 au, peaking at $\sim5$\,au, shows the range in which viscous heating in an optically thick disc dominates.
When the binary separation is wider and the cavity and disc scale increase, the lower densities reduce viscous heating and the effective opacity, which makes cooling more efficient. 
The flaring angle increases and irradiative heating becomes the dominant heat source for systems around binaries with more than $\sim 10$ au in cavity size. 
In this theoretical model, in-plane diffusion of heat aids the thermal transport and leads to $\sim5\times$ faster cooling compared to models with surface cooling alone (see green and orange curves in Fig.~\ref{fig:1d_beta}).

In the context of a planetary companion, \cite{2020Miranda_I} showed that cooling in the regime of the orbital timescale ($\beta=t_\text{cool}\OmegaK\sim 1$) results in the formation of a deep, narrow cavity around the planet. For binary systems, the results of \cite{2022Sudarshan} showed that this cooling regime also leads to smaller and circular binary cavities. The radial dependence of the cooling time demonstrated in Fig.~\ref{fig:1d_beta} and the sensitivity to the cavity size and shape to the cooling timescale indicate that cavity properties should differ around binary systems on different radial scales.

On 1~au scales, the system is predominantly optically thick and the efficiently-cooling range with $\beta\sim1$ is confined to the narrow cavity edge, which renders cooling around the cavity largely inefficient. On the other hand, for the largest binary systems with $\ab =25$~au the disc is optically thin and cools rapidly, especially when considering the effects of in-plane radiative cooling. In the intermediate regime with a binary separation $\ab=5$~au, the entire disc cools at $\beta\sim1$. Based on the above, we therefore expect smaller, circular cavities for $\ab=5$\,au, and larger, more eccentric configurations otherwise.


%
\section{Cavity shapes and sizes at different radial scales} \label{sec:results-shape}

\begin{figure}
	\centering
	\resizebox{\hsize}{!}{\includegraphics{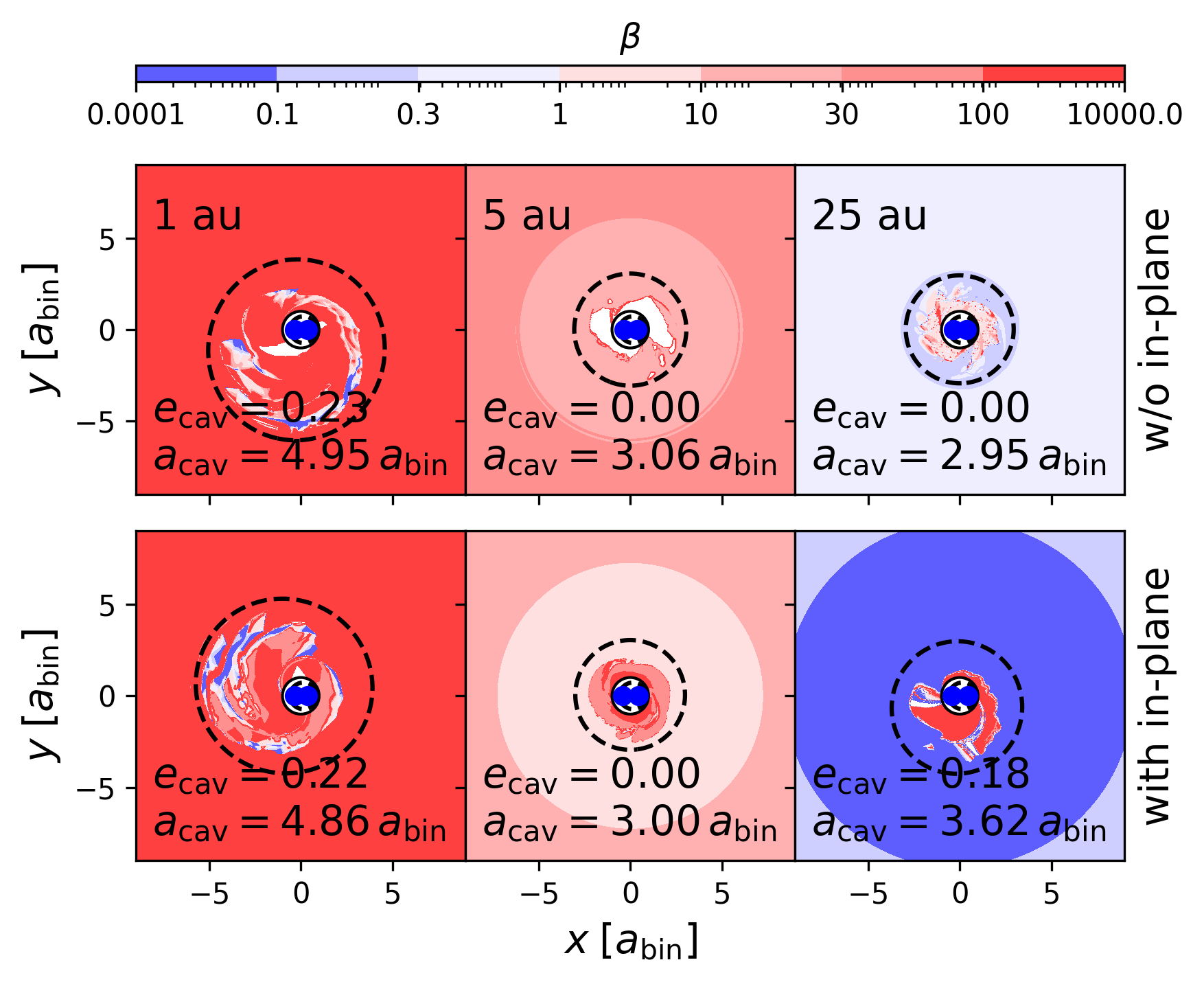}}
	\caption{2D cooling timescale map. The maps show  a snapshot after 29~k$\Tb$ evolution. The colour denotes the orbital cooling time scale $\beta$. The white dashed lines marks the location of the inner cavity. The upper models include all heating and cooling terms with only surface cooling, the lower models in addition include in-plane cooling. The binary orbit has $\eb=0.3$ in all models, and $\ab=1~$au in the left panel, $\ab=5~$au in the middle panel and $\ab=25~$au in the right panel.}
	\label{fig:2d_beta}
\end{figure}

We can compare the 1D prediction for the cooling time to the local cooling in 2D disc models using Eq.~\ref{eq:beta-cooling}. Fig.~\ref{fig:2d_beta} shows the cooling profile of discs with $\eb=0.3$. In that figure and all subsequent heatmaps of quantities in the $\{R,\phi\}$ plane around the binary, dashed ellipses mark a simple fit to the cavity size following \citet{2022Sudarshan}: 
the angular position of the apocentre is linked to the highest density in the disc where the gas within its elliptic orbits is slowest. By finding the density maximum on the opposite side on the line of the apocentre and the centre of the binary as focal point we find pericentre. From these points we define the cavity as the radial position where the density drops to $\leq10\%$ of the maximium density in peri- and apocentre.

In the case of heating and cooling without the in-plane cooling, the reduced heat transport pushes the cooling time in the disc generally to $\beta>0.3$ and to higher values if viscous heating is significant. Otherwise, the increasing irradiation angle of the disc will slowly increase the heating for larger distances.
The thermal state of the cavity (inside the black dashed line) is generally different from the disc. The gas spirals in the inner cavity  are cooling more quickly, while the empty regions in the disc appear strongly heated. In discs that are more circular, the remaining gas in the cavity is less concentrated than in the well-defined spiral wakes in discs that are excited. 

When in-plane cooling is included, heat can be transported away more efficiently, which reduces the magnitude of cooling timescale by a factor of $\sim5$. While this is insignificant for the close viscously heated discs, it makes a notable difference for the $\ab=5$~au scale. However, the disc still remains near the orbital cooling time. The most relevant difference happens for the largest model, where the diffusion of the irradiation heat through the optical thin regime helps the disc to cool much more efficiently as densities drop more quickly than the temperature within the disc and the opacity reaches $\kappa<1$ at the disc temperatures of $T<70~$K. Thereby any heating from the spirals in the disc dissipates quickly and the overall cooling time in the disc drops significantly to $\beta<0.3$. In the cavity, the waves cause heating following the spirals in the nearly empty regions.

This change in cooling efficiency change the wave propagation and by extension the disc shapes for these different scale regimes.

\subsection{Close binaries} \label{sec:1au}

\begin{figure}
		\centering
		\includegraphics[width=\columnwidth]{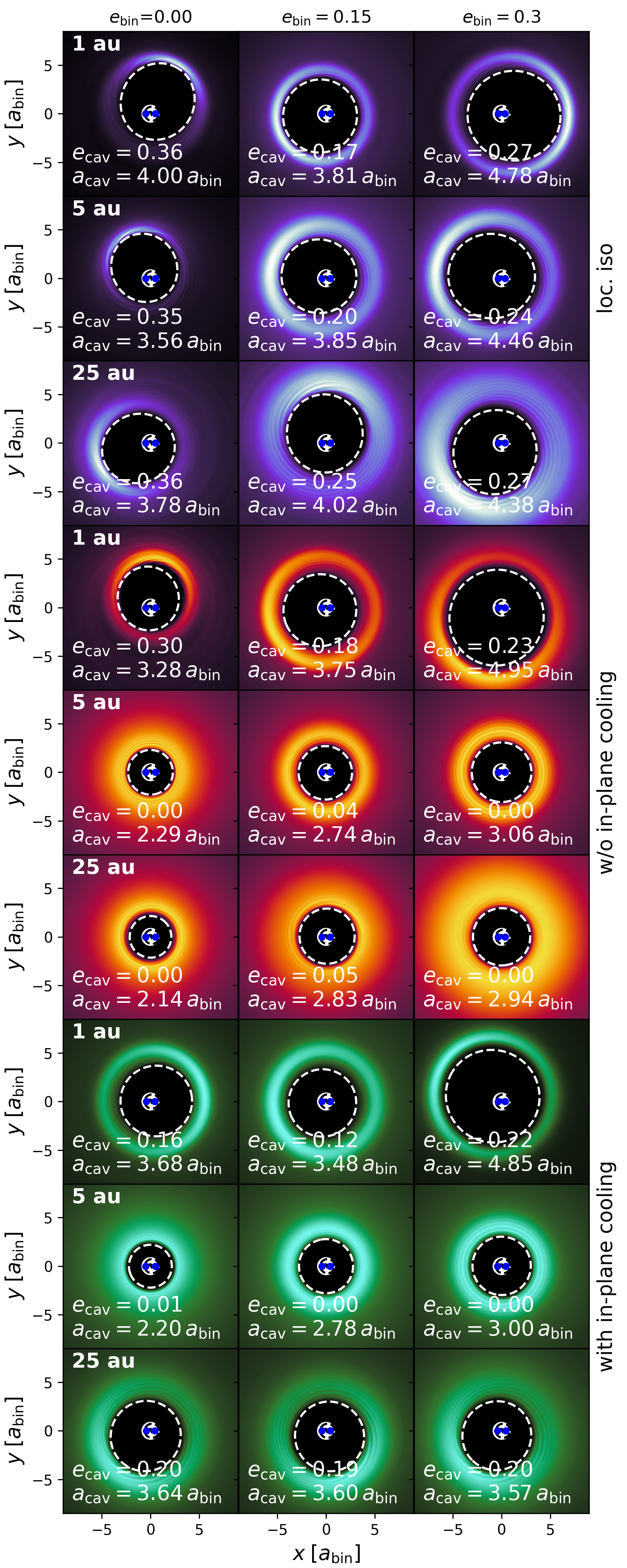}
		\caption{Two-dimensional surface density maps for different thermodynamical prescriptions (different coloured sets), $\ab=[1,5,25]~$au (top to bottom in each set) and $\eb=[0,0.15,0.3]$ (left to right). The cavity edge is marked by a white dashed line.}
		\label{fig:1au_2d}
	\end{figure}

The disc shapes for $\ab=1~$au in Figure~\ref{fig:1au_2d} all show large eccentric cavities independent of their thermodynamical prescription. As shown in previous work \citep[e.g.][]{2017Thun, 2020Ragusa, 2024Penzlin}, the disc are most circular in a binary system with $\eb\approx0.15$, and increase in size and eccentricity for larger $\eb$, while the cavity eccentricity reaches also high values for near circular binary configurations ($\eb\approx0$) except for the in-plane cooling case. One difference between the isothermal disc with $h_0=0.016$ and models including thermal evolution is the width of the inner disc maximum. As the viscous heating through the spirals in the disc gas locally increases the temperature and pressure, the maximum gas density becomes slightly puffed up and wider than in the isothermal case.

\begin{figure}
		\centering
        \includegraphics[width=\textwidth]{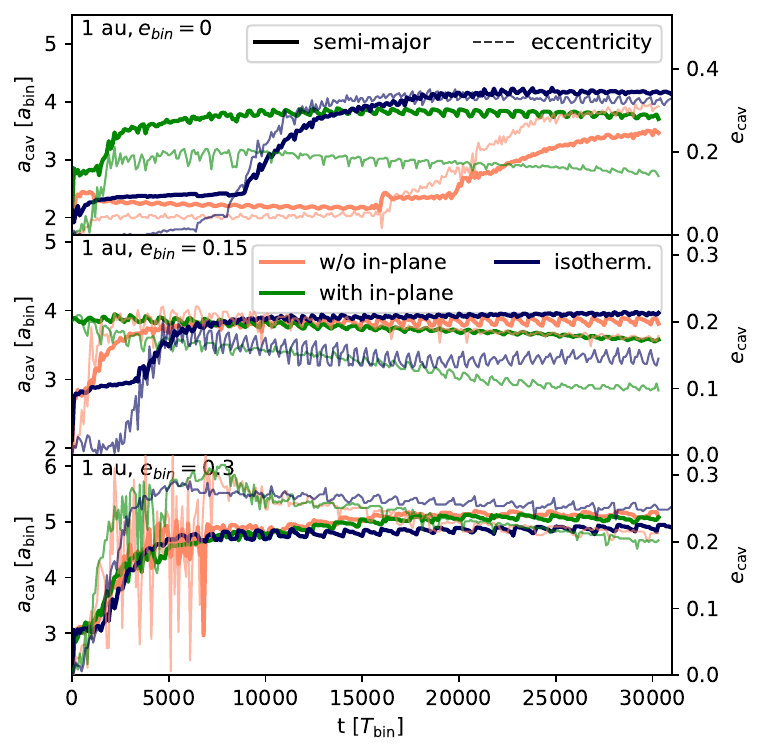}
		\caption{Evolution of the cavity size and eccentricity for models with $\ab=1~$au. Thick lines mark the cavity semi-major axis, thin lines mark the cavity eccentricity. The dark blue models are locally isothermal, the green models include in-plane cooling and the light red models do not include in-plane cooling. The binary eccentricity of all  model in a panel is from top to bottom $\eb=[0,0.15,0.3]$.}
		\label{fig:evo1}
	\end{figure}

To ensure that the circumbinary disc all had enough time to become eccentric, all model reached at least $3 \times 10^4~\Tb$. 
The evolution of the cavity shape for the 1~au case is shown in Fig. \ref{fig:evo1}.

For the systems with a binary separation of 1~au all disc models reach a similar final state, with a cavity size between 3.5 and 5 and an eccentricity between 0.1 and 0.35, depending on the binary eccentricity in the model.

One notable detail in the simulations is that the initial excitation can have an arbitrary delay, especially in the case of the near symmetric potential of the $\eb=10^{-2}$ as the stellar masses are equal. However, as soon as an asymmetry occurs in the disc it quickly grows.
Even though the viscosity is low with an equilibrium scale height of 1.6\% and $\alpha=10^{-4}$ the cavity size and eccentricity rises within 10k~$\Tb$ to $>90\%$ of its value at the end of the simulation. This is much faster than the viscous timescale even in the inner system.

\subsection{Binaries with intermediate separations} \label{sec:5au}

The systems with a binary separation of 5~au demonstrate the stark difference that the cooling time can cause. 
The locally isothermal models achieve the very large and eccentric cavities as the local scale height reaches $h \sim 0.04$ near the disc edge \citep{2024Penzlin}. The correlated pressure increase leads to wider density maxima than in the $\ab=1$~au case in the isothermal models.

In contrast, the shape of the thermally evolving disc in Fig.~\ref{fig:1au_2d} is very small and circular with small cavity sizes that are consistent with the instability limits of circumbinary orbits \citep{1999Holman} or the  short-evolution, circumbinary models by \cite{1994Arty}. Also, the over-density ring near the inner edge is thinner as the wave propagation is restricted.

\begin{figure}
		\centering
        \includegraphics[width=1\columnwidth]{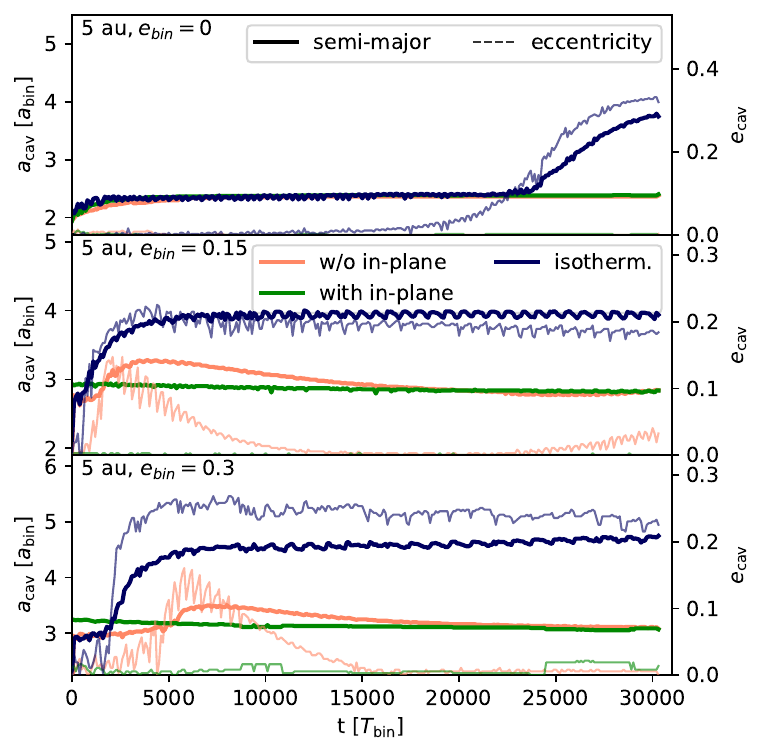}
		\caption{Evolution of the cavity size and eccentricity for models with $\ab=5~$au, similar to Fig.~\ref{fig:evo1}.}
		\label{fig:evo5}
	\end{figure}

This is different from the meta-stable state that we could observe in the $\ab=1~$au before a sufficient perturbation occurred, especially for near circular binaries. For $\ab=5~$au, Fig. \ref{fig:evo5} shows that the there is no growth of significant eccentricity through the whole non-isothermal simulations of 30k~$\Tb$.
While the locally isothermal discs grow in cavity size and eccentricity to $\acav>4$ and $\ecav>0.25$ the model that include heating and cooling, both remain circular after overcoming initial conditions. Any excitation that occurs damps back down within 10k~$\Tb$.
This happens because of the wave damping thermal conditions between the isothermal and the adiabatic limit laid out in \cite{2020Miranda_I}.

 
\subsection{Wide binary discs} \label{sec:25au}

The case of wide binaries, like GG~Tau or HD~142527, shows the role that the in-plane cooling can play in changing the shape of the binary cavity and disc. In Fig.~\ref{fig:1au_2d}, we can see that with in-plane cooling the disc becomes comparable to the locally isothermal case, meaning that the cooling timescale allows effective wave propagation. However, when the thermal evolution only considers vertical direction in the case without inplane cooling the disc can not cool effectively enough and remains in a wave damping regime, which lead to similarly small and circular cavities as in the $\ab=5~$au case in Fig.~\ref{fig:1au_2d}.

\begin{figure}
		\centering
        \includegraphics[width=1\columnwidth]{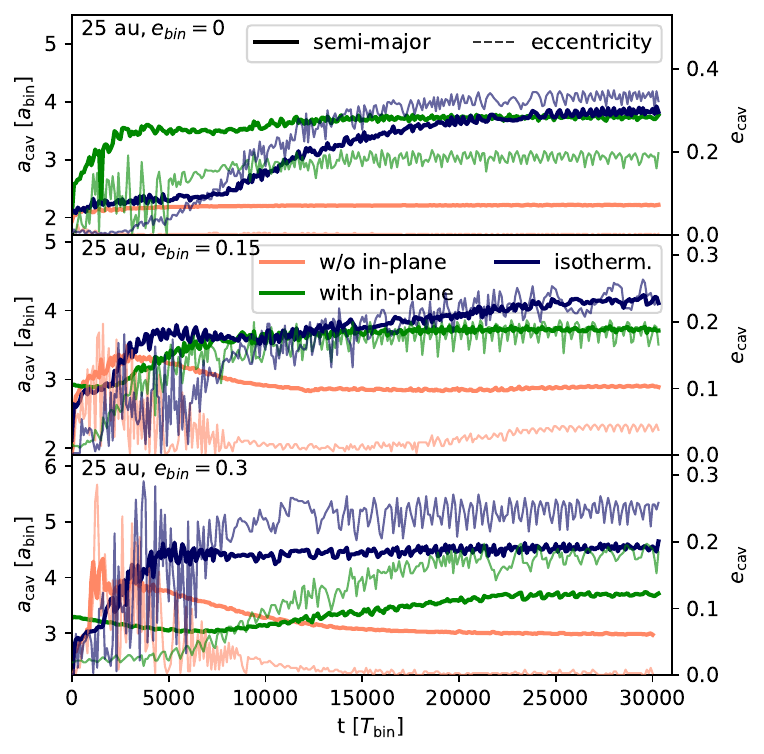}
		\caption{Evolution of the cavity size and eccentricity for models with $\ab=25~$au, similar to Fig.~\ref{fig:evo1}.}
		\label{fig:evo25}
	\end{figure}

To ensure that this is not just a meta-stable state arising from a symmetric initial condition, Fig.~\ref{fig:evo25} shows that the system without in-plane cooling is never able to reach any significant excitation within the simulation time. Meanwhile, the models with the in-plane cooling term can experience some variability but still re-excite to a convergent state with a large eccentric cavity. 
The high pressure in the case of this very large aspect ratio with $h\sim0.1$ leads to overall smaller cavities compared to the $\ab=1~$au case. 
The aspect ratio changes the shape of the disc as seen in the locally isothermal case leading to smaller cavities for the high specific pressures in the largest discs. However, the change in cooling time, also affects the structure of the disc significantly, especially the model with higher eccentricity ($\eb=0.3$) and $\ab=25$~au. As the cooling time scale within the disc remains finite, some level of wave damping is expected above the level of the fully excited locally isothermal disc seen in other simulations \citep[e.g.][]{2020Ragusa,2024Penzlin}.

 
\section{Synthetic observations}\label{sec:synthetic}

As the different binary separation apply for different systems, we can compare the resulting observation for such systems. Since our simulations are only gas, they are best to compare to scattered-light and CO observation of systems.
We use radmc3d \cite{radmc} to created synthetic observation based on the models which include in-plane cooling.
A symmetric Gaussian convolution filter of 30 milliarcsecs is applied to the image.

To produce the synthetic $\rm ^{12}CO(3-2)$ data cubes from the PLUTO simulation of a disc around a binary with 5 au separation, we followed the procedure presented in \citealt{2024Barraza} (see further details and references therein). However, we used directly the simulation gas temperature as input into radmc3d.

To expand the 2D numerical simulation to 3D we assuming the vertical density structure follows $\rho(z) = \Sigma (2\pi H^2)^{-0.5}\exp(-z^2/2H^2)$. The optical depth $\tau$ is then proportional to the integration from infinity to $z$ of this Gaussian profile, which we scale up linearly to reach a depth of 100 at the mid plane. We assume that the temperature will reduce below $\tau=1$ and decay to 0.1 of the isothermal temperature $T_\mathrm{iso}$ as calculated in Eq.~\ref{eq:cs} and, hence, be given by $T(z)=T_\mathrm{iso} (1 - 0.9 \tau (1+ \tau)^{-1})$. The vertical temperature structure created by this method is an approximation that likely underestimates the true temperature, however, it is sufficient to create a optical surface necessary for the synthetic CO maps.

We compute data cubes centred at $\sim 345.796 \rm\, GHz$, with channels covering $\rm \pm 8.875 \,km \,s^{-1}$ and a velocity resolution of $\rm 0.25 \,km \,s^{-1}$, reached by binning 10 channels with finer $\rm 0.025 \,km \,s^{-1}$ spacings. In addition, we included a $\rm \pm 25 \,m \,s^{-1}$ micro-turbulent broadening to prevent artifacts in the line ray-tracing at the cold outer regions. For the Cs Cha-comparison, we assumed the system parameters as presented in \citealt{2022Kurtovic}: a distance to the source of 169 pc, a combined mass star of the binaries of $1.9M_{\odot}$, a disc inclination of 17.86 deg, and PA of 263.1 deg.

To create the dust observation we first continued the simulations using \texttt{Pluto} \text{v.4.4}, which includes support for dust as a pressureless fluid \citep[see][for details]{2024dust} for $2000~\Tb$. The dust fluid component corresponded to grains with a bulk density of $\tilde{\rho} = 2.08\,\text{g}/\text{cm}^3$ and a grain size of 1\,mm, and was initialised with a dust density profile of $\Sigma_\text{d}=0.01\,\Sigma_\text{g}$ exterior to the pressure maximum at the cavity edge, and $\Sigma_\text{d}=10^{-6}\,\Sigma_\text{g}$ otherwise. Dust back-reaction on the gas was enabled as well, which has been shown to result in more circular cavities \citep{2022Coleman}, although we note that for the dust-to-gas ratio we employed this circularisation effect should be minuscule.

\begin{table}[]
\begin{tabular}{|c|c|c|c|c|}
\hline 
parameter & CsCha [i] & sim [i] & GG Tau A [ii] & sim [ii]\\ 
\hline 
$M_\mathrm{bin} [M_{\odot}]$ & $2\pm0.2^1$ &1.9 & $1.1 - 1.6^2$ & 1.2\\ 
$M_2/M_1$ & $0.66-1^1$ & 1 & $0.66 - 1^2$ & 0.7\\ 
$\ab$ [au] & $5.0\pm 0.2^1$ & 5 & $34 \substack{+5.9 \\ -2.8}^3$& 39.9\\ 
$\eb$ & $0.40 \pm 0.04^1$ & 0.3 & $0.28 \substack{+0.05 \\ -0.14}^3$ & 0.28\\ 
\hline 
\end{tabular}
\caption{Binary parameters for Cs Cha [i] and GG Tau [ii] as stated in the literature (row 2, 4) and used in the simulation (row 3, 5). $M_\mathrm{bin}$ is the total binary mass in solar masses, $\ab$ and $\eb$ are the semimajor axis and eccentricity of the binary orbit and the binary mass ratio is given by $M_2/M_1$. 
$1:$ \protect{\citet{2024Ginski}}, $2:$ \protect{\citet{2020Keppler}}, $3:$ \protect{\citet{2011Koehler}}
}
\label{tab:obs}
\end{table}

For the system parameters we used the measurement in \cite{2022Kurtovic} for CsCha and \cite{2020Keppler} and \cite{2011Koehler} for GG Tau as summaries in Table~\ref{tab:obs}. 
More recent orbital parameters for GG Tau can also be found in \cite{2024Toci}.
We then processed the resulting dust and gas density maps through \texttt{RADMC-3D} \citep{radmc}, assuming a mixing--settling equilibrium for the vertical direction with $\alpha=10^{-4}$ following \citet{fromang-nelson-2009}. The dust component of the \texttt{Pluto} model was used as is, and the gas component was translated to a population of small grains by multiplying $\Sigma_\text{g}$ by 0.01 and assigning a grain size of 0.1\,$\mu$m. The dust opacities were then computing for $0.1\,\mu$m and 1\,mm grains using \texttt{OpTool} \citep{optool} following the \texttt{DIANA} standard \citep{woitke-etal-2016}. We computed the thermal structure of the disc using the \texttt{mctherm} command of \texttt{RADMC-3D} with $10^9$ photons assuming anisotropic scattering using the approach of \citet{henyey-greenstein-1941} with the anisotropy factor computed via \texttt{OpTool}. Finally, we used the \texttt{image} command to generate images of thermal continuum emission at 0.9\,mm and scattered light emission at $1\,\mu$m.

\begin{figure}
		\centering
        \includegraphics[width=1\columnwidth]{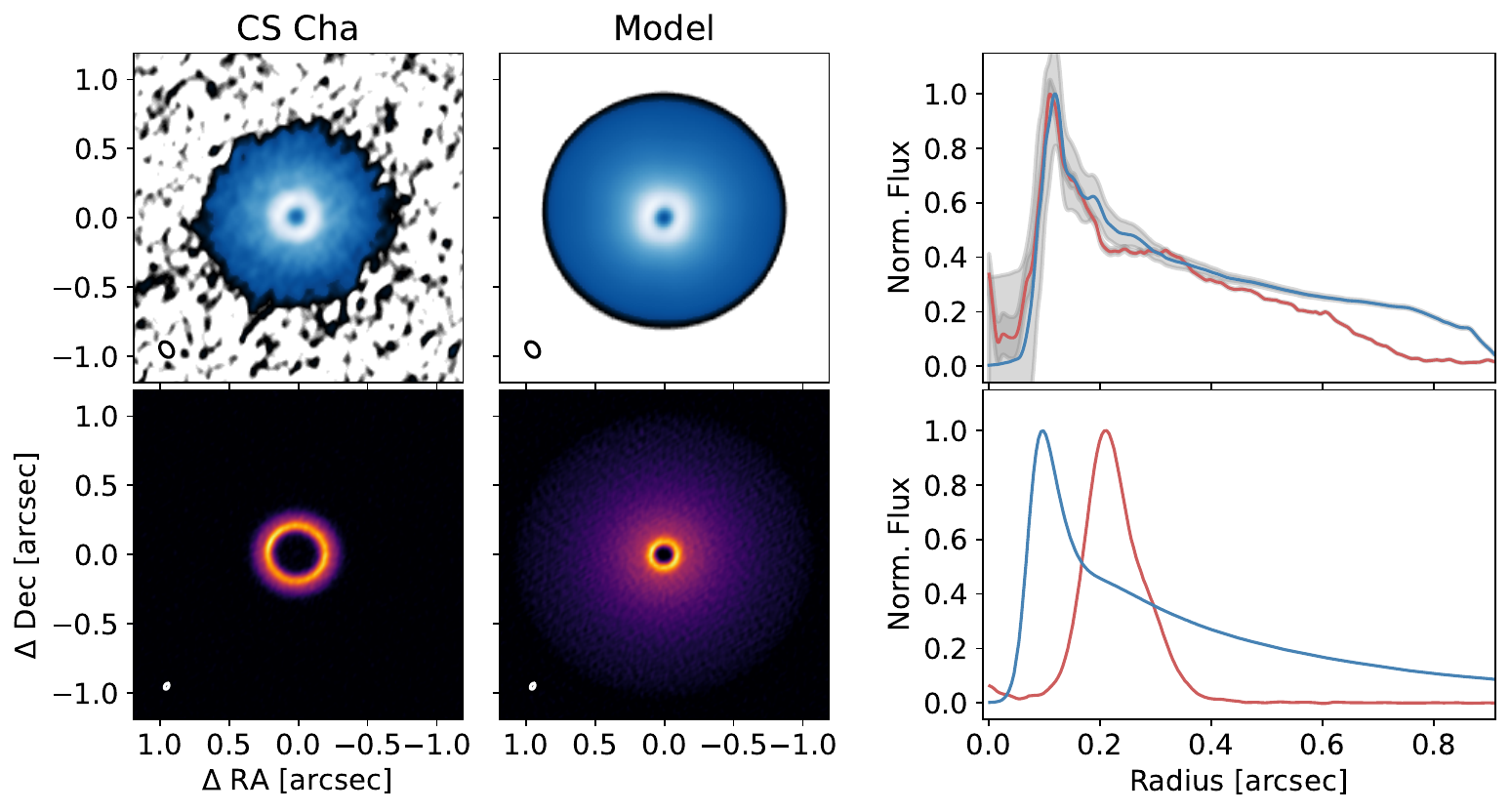}
		\caption{Synthetic observation (middle) compared to observation of Cs~Cha \citep{2022Kurtovic} (left). 
        $870~\mu$m dust continuum emission from the observation (bottom left) and the continued dust simulation using radmc3d (bottom middle). 
        $\rm ^{12}CO J:3-2$  integrated intensity maps (zeroth moment) from synthetic predictions (top middle) and ALMA observations of Cs~Cha (top left). The synthetic images are normalized to the maximum emission. The symbols in the bottom left of each panel indicate the convolution size. The normalized, azimuthally averaged intensity profile for the dust (red lines) and the CO (blue lines) are compared between the synthetic model(top right and the observation(bottom right).}
		\label{fig:com_CsCha}
	\end{figure}

The Cs-Cha system is an example of a disc with a circular cavity.
The completely circular shape of this system, could so far only be explained by models that included the interference of a planet between the binary wakes and the disc as in \cite{2022Kurtovic}. However, wave damping through thermodynamics can be as effective to circularise the disc as is show in the comparison to the $\rm ^{12}CO(3-2)$ ALMA observation \citep{2022Kurtovic} in Fig.~\ref{fig:com_CsCha}.
Assuming a binary separation of $\ab=5~$au and the binary at a distance of 169~parsec we can recover the shape of the CsCha observation gas. 
For the dust simulations, we might be missing important aspect of the dust evolution like the fragmentation near the inner edge of the disc. The inner most region even of the circular disc will still reach higher turbulent velocities due to the proximity to the binary which can inhibit dust grow \citep{2021Pierens}.
As the semi-major axis of the binary is uncertain and potentially wider than 5~au, the size of the synthetic observations does not match the AMLA images perfectly due to our specific choice of binary orbital elements. However, the CO observations is well within the expected range allowed by the uncertainty.

\begin{figure*}
		\centering
        \includegraphics[width=1\columnwidth]{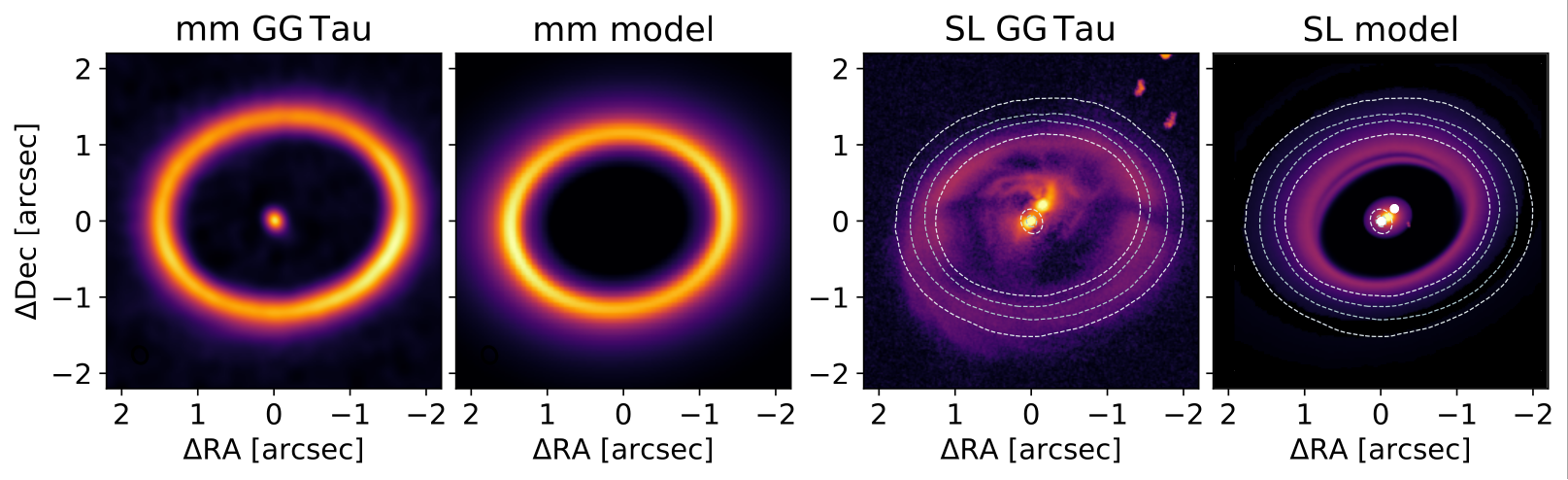}
		\caption{
        The continuum emission \citep[][left panels]{2024Rota} and scattered light image \citep[][right panels]{2020Keppler} of GG Tau A compared to the model. The continuum is added as dashed contours for reference to the scattered light images (SL). The white dots in the right panel mark the location of the binary components in the simulation.}
		\label{fig:com_GGtau}
	\end{figure*}

We can also compare the wide models listed in Table~\ref{tab:obs} to the GG~Tau system, following the same steps as described before.
In Fig.~\ref{fig:com_GGtau}, the left two panels compare the continuum emission between observation \citep{2024Rota} and simulation. Both show a strong circular trapping of millimetre dust beyond the disc inner cavity position at the density maximum of the disc. We compare also the scattered light observation of the GG~Tau system \citep{2020Keppler} with our simulation assuming vertically settled dust following the gas surface density with a grain size of $870~\mu$m. To emulate the infalling material near the binary, we fill the region inward of our domain with $5 \times 10^{-4}$ of the peak density with a aspect ratio of 15\% in a hydrostatic vertical profile for the scattered light model.
This smooths out and removes emission from the inner disc as already investigated in \cite{2019Brauer}.
The exact size of the cavity will also depend on the precise cooling time scale in addition to the binary parameters. The cooling depends on an uncertain dust composition, disc mass and turbulence level.
We recover a dynamic behaviour in our model that matches the structures in GG Tau.

The eccentric scattered light image is off-centre compared to the continuum emission. This is caused by the radial decay in eccentricity and decay in the related azimuthal density differences in the disc between cavity wall and density maximum. Spirals appear at the upper surface of the scattered light image in both model and observation. However, the instreaming material in the observation toward the binary cannot be captured by a planar model that is symmetrically stratified, and these streams above the midplane would need full 3D hydrodynamic modelling.
We discuss the features in the disc and the exact eccentricity of the millimetre and micrometre dust positions in the disc further in Appendix~\ref{sec:appA}.
The thermal conditions lead to some but not a complete reduction in the eccentricity. This resolves the tension of too high eccentricities reached in previous locally isothermal simulations in \cite{2024Toci} to match observations. 

\section{Discussion}\label{sec:discuss}

The 2D simulation in this work require assumptions and approximations that allow us to study the circumbinary disc evolution through long-term hydrodynamic simulation. This concerns the time evolution of disc and binary, viscous dynamics in the 2D disc model, the thermodynamic model and additional dynamic components to the systems like dust dynamics or planet interaction with the disc.

\subsection{Evolution of disc and binary}

All simulations have run for $\geq 3\times 10^4~\Tb$ and this evolution time is important, particularly for the binary setup we chose with equal mass or near equal mass binaries. Such binaries can remain in a symmetric state for even $15~\mathrm{k}\Tb$ for circular binary orbits \cite{2020Keppler}. This is a remnant of unphysically symmetric setup conditions, especially when the binary orbit is not evolving with the disc to maintain similar orbits through the simulation.
For this reason the \cite{2024KITP} comparison study applies already asymmetric initial condition in the disc to spark excitation for all the equal mass systems, nevertheless the eccentricity growth shows some orbits delay between different setups.
In our simulation we initialized all disc with a minimal eccentricity of 0.01, this appears to be less effective than perturbing the disc to start of the growth of eccentricity as the low eccentricity disc at 1 and 5~au remain circular for an extend initial phase. However,
if discs have the conditions to become eccentric they will not return to a symmetric state once excited. The convergent state of the disc and the inner cavity is eccentric and precessing \citep[e.g.][]{2017Thun, 2019Munoz, 2020Ragusa, 2024Penzlin}.
The binary in our model is not included in the domain and does not evolve its orbit due to the gravitational or accretion interaction with the disc. 

Many locally isothermal studies show that the binaries' mutual orbit shrinks due to the interaction with the disc \citep{2020Tiede, 2021Dittmann, 2023siwek}. For example, \cite{2020Tiede} shows that this gravitational torque dominates and leads to an in-spiralling of the binary orbits. In the model with a cooling times that leads to eccentricity damping, the gravitational torque between planets and binaries in more circular discs may be weakened. To understand if this reduction in torque is sufficient to allow the momentum advected onto the binaries to expand the orbits instead will be part of future investigations.

\subsection{Other methods for circularisation}

In this work, we do not include the dust evolution. \cite{2022Coleman} has investigated how dust can also aid the circularization of circumbinary disc. The shorter 3D simulations by \cite{2021Pierens, 2023Pierens} also show a reduced eccentricity of the disc, {as the momentum can propagate in the vertical direction rather than excite the eccentricity in full.}

Hence, our gas disc represents an upper limit on the expected eccentricity of the disc.

\cite{2022Kurtovic} has shown that the Cs~Cha disc is consistent with a planet circularizing the cavity. While the circularization can be caused by the thermodynamic interaction it can also be caused and maintained by planets \citep{2021Penzlin}. \cite{2023Follette} reported some evidence for a companion "c" close to the inner disc of the Cs Cha binary. Given the current uncertainties in the orbital parameters of Cs Cha, it is difficult do separate the scenarios. However, the thermally truncated cavity is up to $1 \ab$ smaller depending on the unknown eccentricity and thereby constraining the orbit closer or additional high resolution kinematic features like planet kinks \cite{2018Teague} or eccentric orbital deviations \citep{2024vel} through the binary could help to distinguish the cases.
While these causes can not be distinguished through the gas density map, they would lead to different signatures in the velocities of the disc \citep{2024vel}. Even a circular circumbinary disc will show weak large-scale eccentric modes due to the wide binary wakes. Meanwhile, a disc with a planet has more local velocity perturbation which are more narrow. We will investigate these difference in a future study.

\subsection{Viscosity and instability in the disc}

To reach simulations times that show whether or not the disc gets excited is only feasible in 2D. Thus, we can not account for 3D effects that might become relevant, like the effects of the eccentric turbulence seen in the 3D simulations by \cite{2021Pierens} for which we assume a viscous $\alpha=10^{-4}$. This viscosity also determines the amount of viscous heating and is in our model just parameterized to a realistically low value.
Another effect that would require the full 3D model would be the radiative transfer in the disc, as we consider only vertically integrated approximation. 

Through extending the length scale of the disc the overall mass of the discs also increases even using the decreasing reference densities.
In this study, we always consider an untruncated, steady disc that extends between $\sim 3 - 40~\ab$, such that our largest disc would extend out to 1000~au. Such a size is beyond the scale of most disc. In observations larger discs are often found to be more massive \citep[e.g.][]{1999Guilloteau}, however, the masses within the domain do not reflect complete realistic disc but match densities needed for a realistic inner disc and cavity, which would truncate at $\leq 200$~au. 
Simulation by \cite{2017Mutter} have considered gravitational instability for the locally isothermal discs and found that the heaviest disc (10-20 mean mass solar nebular) are able to lead to a small reduction in the cavity size.

\subsection{Effects of radiative diffusion}

While we consider the effects of cooling due to the in-plane transport of a radiative flux (see Eq.~\eqref{eq:beta_rad}), we do not capture the effects of radiative diffusion, which would require incorporating the term in Eq.~\eqref{eq:Q_rad} instead. The effect of this omission can be two-fold.

\citet{ziampras-etal-2023} showed that, while a full treatment of radiative diffusion results in a difference in gap opening efficiency in the context of a planetary companion, this difference can be ``absorbed'' into an effective reduction of the cooling timescale by a factor of 1.5--2 depending on the optical depth. Given that the cooling timescale spans 4 orders of magnitude in our models (see Fig.~\ref{fig:1d_beta}), such a correction does not affect the quality of our results regarding the size of the cavity.

At the same time, however, the omission of radiative diffusion will affect the thermal structure of the cavity. For example, on small radial scales, shock heating can become important \citep[e.g.][]{2016Rafikov}, and the lack of thermal diffusion around shock fronts can change their heat input into the disc, possibly smoothening temperature peaks along azimuth. Overall, while relevant for a more accurate model of the cavity, radiative diffusion should not affect the orbital properties of the cavity in the context of our results.

\subsection{General applicability of our results}

In our models, we assume a particular temperature, density, and opacity model that results in the radial cooling timescale profile shown in Fig.~\ref{fig:1d_beta}. A different opacity model \citep[e.g.][]{semenov-etal-2003,woitke-etal-2016,birnstiel-etal-2018} or dust size distribution due to dust growth \citep[e.g.][]{birnstiel-2024} would certainly influence the radial range where $\beta\sim1$ and therefore the scale where circular cavities are expected. The same is true for different choices in disc mass and/or stellar parameters. However, the radial pressure-density structure and the radially dependent irradiation, will lead to differences between systems with different physical scales that are otherwise similar. As the discs loses material over time they will get optically thinner very slow creating a small reduction in the local cooling time.
Nevertheless, we expect that for reasonable disc parameters the $\sim$au scale will typically be optically thick and the $\sim$50\,au range optically thin, with the $\beta\sim1$ regime somewhere in the 5--20\,au region. Our results in Sect.~\ref{sec:results-shape} are to be interpreted as a general rule of thumb when modelling circumbinary discs, but the properties of particular systems are subject to existing observational constraints for those systems.

\section{Conclusion}\label{sec:conclusion}

To understand the impact that thermodynamic conditions can have on the structure of the disc, we ran a set of 2D radiation hydrodynamical simulations using the \pluto{}. We varied the binary eccentricity, investigated three size scales $\ab=[1,5,25]$~au, and compared three different thermodynamical models: locally isothermal, radiative with viscous heating, stellar irradiation and surface cooling, and radiative with the addition of in-plane cooling.

The simulations showed that the excitation of the eccentricity and size of the inner cavity is sensitive to the wake propagation caused by the binary star motion.
On one side, the theoretical case of an isothermal disc represents an instant cooling and allows strong excitation of the disc eccentricity and on the other end of theoretical thermodynamic conditions an adiabatic disc represents an never-cooling disc which also allows strong excitation. However, all realistic disc are in the regime of a finite cooling time in between these extreme conditions, which leads to a variation in how pressure waves propagate.
Similar to the planetary scenario explored in \cite{2020Miranda_I, 2020Miranda_II} and demonstrated in a binary context by \citet{2022Sudarshan}, radiative damping inhibits the spiral angular momentum flux for intermediate cooling timescales on the order of the local orbital period. For systems that meet this criterion, the circumbinary cavities remain small and circular.

For disc models with viscous and irradiative heating and cooling via thermal emission, such conditions become relevant if the binary is separated by $\sim 5$~au or circumbinary cavities are $\sim20$ -- $50$~au in size. For even larger discs, in-plane radiative cooling can aid with cooling such that the disc cools efficiently enough to be considered quasi-locally isothermal, exciting large and eccentric cavities.
For the smallest separation of 1\,au, the high optical depth leads to inefficient cooling and quasi-adiabatic conditions within the disc, allowing larger eccentric cavities to form.

This size scale dependent behaviour of the circumbinary disc can explain the level of eccentricity in the observations of Cs~Cha and GG~Tau with one physical model.

\section*{Acknowledgements}

AP \& AZ would like to thank Kees Dullemond, Richard Nelson, Richard Booth and James Owen for discussions, advice and encouragement. 
The authors also thank Miriam Keppler and Alessia Rota for sharing and advice on their GG Tau observations. 
AP acknowledges support from the Royal Society in the form of a University Research Fellowship and Enhanced Expenses Award. 
AZ acknowledges support by STFC grant ST/P000592/1, and AP~\&~AZ acknowledge support from the European Union under the European Union's Horizon Europe Research and Innovation Programme 101124282 (EARLYBIRD). 
This work was supported by an MIT-Imperial Seed fund provided through MISTI. PP acknowledges funding from the UK Research and Innovation (UKRI) under the UK government’s Horizon Europe funding guarantee from ERC (under grant agreement No 101076489). 
The simulations were performed with the support of the High Performance and Cloud Computing Group at the Zentrum f\"ur Datenverarbeitung of the University of T\"ubingen, the state of Baden-W\"urttemberg through bwHPC and the German Research Foundation (DFG) through grant no INST 37/935-1 FUGG, and Cambridge Service for Data Driven Discovery (CSD3), part of which is operated by the University of Cambridge Research Computing on behalf of the STFC DiRAC HPC Facility (www.dirac.ac.uk). The DiRAC component of CSD3 was funded by BEIS capital funding via STFC capital grants ST/P002307/1 and ST/R002452/1 and STFC operations grant ST/R00689X/1. DiRAC is part of the National e-Infrastructure. This research utilized Queen Mary's Apocrita HPC facility, supported by QMUL Research-IT (http://doi.org/10.5281/zenodo.438045).  
NTK has been funded by the Deutsche Forschungsgemeinschaft (DFG, German Research Foundation) - 325594231, FOR 2634/2. 
MBA acknowledges the MIT SuperCloud and Lincoln Laboratory Supercomputing Center for providing HPC resources that have contributed to the research results reported within this paper.
Views and opinions expressed are those of the authors only and do not necessarily reflect those of the European Union or the European Research Council. Neither the European Union nor the granting authority can be held responsible for them.

This paper makes use of the following ALMA data: ADS/JAO.ALMA\#2017.1.00969.S, ADS/JAO.ALMA\#2018.1.00532.S. ALMA is a partnership of ESO (representing its member states), NSF (USA) and NINS (Japan), together with NRC (Canada), MOST and ASIAA (Taiwan), and KASI (Republic of Korea), in cooperation with the Republic of Chile. The Joint ALMA Observatory is operated by ESO, AUI/NRAO and NAOJ. 

\section*{Data Availability}
The data underlying this article will be shared on reasonable request to the corresponding author. The observations are publicly available through ALMA and ESO archives. 

\bibliography{paper}
\bibliographystyle{mnras}

\begin{appendix}

\section{Dynamic eccentricity in GG Tau}\label{sec:appA}

\begin{figure}
		\centering
        \includegraphics[width=1\columnwidth]{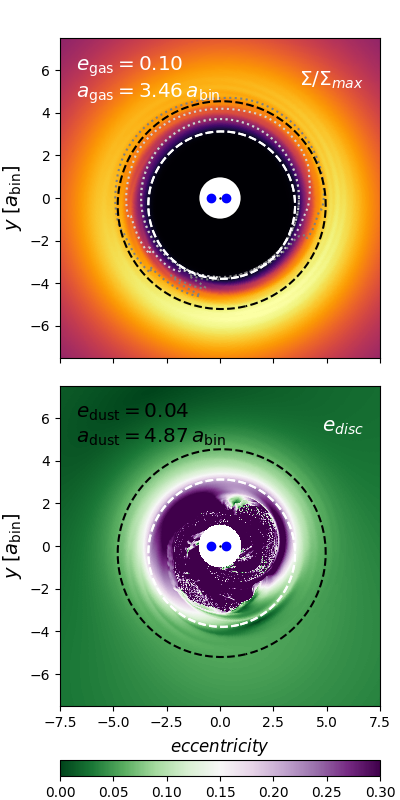}
		\caption{Surface density (top) and eccentricity (bottom) of the GG Tau disc simulation. The white dashed line indicates the inner gas cavity, the black dashed line indicates the dust inner cavity. The light grey and dark grey dotted line mark a eccentricity of 0.08 and 0.06 in the top panel. The dots show the binary locations.}
		\label{fig:sim_GGtau}
	\end{figure}

To understand the different behaviour of the circular dust continuum and the less symmetric scattered light observation with its notable spiral features it helps to look a the dynamic eccentricity of in the simulation. The dynamic eccentricity uses the velocities vector in for each cell to calculate the orbital parameters of an eccentric Keplerian orbit around the centre of mass \citep[see also ][ for more details]{2024Penzlin}.
Figure~\ref{fig:sim_GGtau} shows the dynamic eccentricity (bottom panel) compared to the disc surface density profile (top panel). The eccentricity at the cavity edge reaches $\sim0.1$. However, within the disc the eccentricity decays quickly and drops to values $\sim0.05$ around the inner edge of the dust density. The decay in the eccentricity depends on the conditions in the disc \citep{2024Penzlin} and the wave damping of the finite cooling time in the disc aids the decay of eccentricity within the disc.
The initial eccentricity in the gas at the cavity location is enough to cause asymmetries between apo- and pericentre and drive notable spirals. However at the location where the dust is trapped the system is less dynamic and the dust can create a unperturbed circular ring in the mid-plane.

\end{appendix}

\bsp	
\label{lastpage}
\end{document}